\documentclass[12pt]{article}
\def\Een{{\bf 1}}
\def\im{{\bf i}}
\def\X{{\bf X}}

\def\E{{\bf E}}
\def\A{{\bf A}}
\def\B{\overline{\bf B}}
\def\C{\underline{\bf C}}
\def\q{\overline{\bf q}}
\def\p{\underline{\bf p}}
\def\F{{\bf F}}
\def\G{{\bf G}}
\def\Z{{\bf Z}}
\def\dif{{\bf d}}
\def\H{{\cal H}}
\def\J{{\cal J}}

\def\L{{\cal L}}
\def\LG{{\cal G}}
\def\M{{\cal M}}
\def\P{{\cal P}}
\def\Q{{\cal Q}}
\def\V{{\cal V}}
\def\T{{\cal T}}

\def\ksi{\overline{\xi}}
\def\pie{\underline{\pi}}

\def\be{\begin{equation}}
\def\ee{\end{equation}}
\def\eea{\end{eqnarray}}
\def\bea{\begin{eqnarray}}
\def\bean{\begin{eqnarray*}}
\def\eean{\end{eqnarray*}}
\title{Noncommutative Configuration Space.\\
 Classical and Quantum Mechanical Aspects\thanks{Extended version of a communication at the XXVth National Meeting of Particle Physics and Field Theory, august 2004, Caxamb\'{u}, MG, Brazil.}}
\author{F.J. Vanhecke\thanks{vanhecke@if.ufrj.br}, C. Sigaud and A.R. da Silva \\ Instituto de F\'{\i}sica, Instituto de Matem\'{a}tica,\\ UFRJ, Rio de Janeiro, Brazil } 
\date{}
\begin{document}
\maketitle
%
%
\begin{abstract}
In this work we examine noncommutativity of position coordinates in classical symplectic mechanics and its quantisation. In coordinates $\{q^i,p_k\}$ the canonical symplectic two-form 
is $\omega_0=dq^i\wedge dp_i$. It is well known in symplectic mechanics {\bf\cite{Souriau,Abraham,Guillemin}} that the interaction of a charged particle with a magnetic field can be described in a Hamiltonian formalism without a choice of a potential. This is done by means of a modified symplectic two-form $\omega=\omega_0-e\F$, where $e$ is the charge and the (time-independent) magnetic field $\F$ is closed: $\dif\F=0$. With this symplectic structure, the canonical momentum variables acquire non-vanishing Poisson brackets: $\{p_k,p_l\}\,=\,e\,F_{kl}(q)$. 
Similarly a closed two-form in $p$-space $\G$ may be introduced. Such a {\it dual magnetic field} $\G$ interacts with the particle's {\it dual charge} $r$. A new modified symplectic two-form $\omega=\omega_0-e\F+r\G$ is then defined. Now, both $p$- and $q$-variables will cease to Poisson commute and upon quantisation they become noncommuting operators. In the particular case of a linear phase space ${\bf R}^{2N}$, it makes sense to consider constant $\F$ and $\G$ fields. It is then possible to define, by a linear transformation, global Darboux coordinates: $\{\xi^i,\pi_k\}=\,{\delta^i}_k$. These can then be quantised in the usual way $[\widehat{\xi}^i,\widehat{\pi}_k]=i\hbar\,{\delta^i}_k$.
The case of a quadratic potential is examined with some detail when $N$ equals 2 and 3. 
\end{abstract}
%
\section{Introduction}\label{inleiding}
The idea to consider non vanishing commutation relations between position operators $[{\bf x},{\bf y}]=i\,\ell^2$, analogous to the canonical commutation relations between position and conjugate momentum $[{\bf x},{\bf p}_x]=i\,\hbar$, is ascribed to Heisenberg, who saw there a possibility to introduce a fundamental lenght $\ell$ which might control the short distance singularities of quantum field theory. However, noncommutativity of coordinates appeared first 
nonrelativistically in the work of Peierls {\bf\cite{Peierls}} on the diamagnetism of conduction electrons. In the limit of a strong magnetic field in the $z$-direction, the gap between Landau levels becomes large and, to leading order, one obtains $[{\bf x},{\bf y}]=i\,\hbar c/eB$. In relativistic quantum mechanics, noncommutativity was first examined in 1947 by Snyder {\bf\cite{Snyder}} and, in the last five years, inspired by string and brane-theory, many papers on field theory in noncommutative spaces appeared in the physics literature. The apparent unitarity problem related to time-space noncommutativity in field theory was studied and solved in {\bf\cite{Doplicher}}. Also (nonrelativistic) quantum mechanics on noncommutative twodimensional spaces has been examined more thorougly in the recent years: {\bf\cite{Duval,Nair,Morariu,Horvathy,Djemai,Berard}}. The above mentionned unitarity problem  
in quantum physics is also examined in Balachandran et al. {\bf \cite{Bal}}.\\
In this work we discuss noncommutativity of configuration space $\Q$ in classical mechanics on the cotangent bundle $T^*(\Q)$ and its canonical quantisation in the most simple case. In section {\bf\ref{Tweede}} we review the classical theory of a non relativistic particle interacting with a time-independent magnetic field $\F=1/2\,F_{ij}(q)\,dq^i\wedge dq^j\;;\;\dif\F=0$. This is done in every textbook introducing a potential in a Lagrangian formalism. The Legendre transformation defines then the Hamiltonian and the canonical symplectic two-form $dq^i\wedge dp_i$ implements the corresponding Hamiltonian vector field. We also recall the less well known procedure of avoiding the introduction of a potential using a modified symplectic structure: $\omega=dq^i\wedge dp_i-e\F$.
The coupling with the charge $e$ is hidden in the symplectic structure and does not show up 
in the Hamiltonian: $H_0(q,p)=\delta^{kl}\,p_k\,p_l/2m\,+\,\V(q)$. In section {\bf\ref{Derde}}, a closed two-form in $p$-space, the {\it dual field}: $\G=1/2\,G^{kl}(p)\,dp_k\wedge dp_l$, is added to the symplectic structure $\omega=dq^i\wedge dp_i-e\F+r\G$, where $r$ is a {\it dual charge}.\\
Such an approach with a modified symplectic structure has been previously considered by Duval and Horvathy{\bf \cite{Duval,Horvathy}} emphasizing  the $N=2$-dimensional case in connection with the quantum Hall effect. We should also mention Plyushchay's interpretation {\bf \cite{PeterMikhail}} of such a dual charge $r$ when $N=2$ as the anyon spin. Considering here an arbitrary number of dimensions $N$, no such interpretation of $r$ is assumed. The crucial point is that, now, both $p$- and $q$-variables cease to Poisson commute and upon quantisation they should become noncommuting operators. In the particular case of a linear phase space ${\bf R}^{2N}$, it makes sense to consider constant $\F$ and $\G$ fields. It is then possible to define global Darboux coordinates with Poisson brackets $\{\xi^i,\pi_k\}=\,{\delta^i}_k$. These can then be quantised uniquely {\bf\cite{John}} in the usual way: $[\widehat{\xi}^i,\widehat{\pi}_k]=i\hbar\,{\delta^i}_k$. However, in general, the dynamics become non-linear and there is no guarantee that the  Hamiltonian vector field is complete. It is then not trivial to quantise the Hamiltonian, which becomes nonlocal. However, for a linear or quadratic Hamiltonian, this is possible and it is seen that the noncommutativity generates a magnetic moment type interaction. The cases $N=2$ and $N=3$ are discussed in detail in section {\bf\ref{Vierde}}. In section {\bf\ref{Vijfde}} we examine the problem of symmetries in the modified symplectic manifold. Finally, in section {\bf\ref{Zesde}} general comments are made and further developments are suggested. In appendix {\bf\ref{mechanica}} we recall basic notions in symplectic geometry and in appendix {\bf\ref{link}} we give a brief account of the Gotay-Nester-Hinds algorithm {\bf\cite{GNH}} for constrained Hamiltonian systems. 
\section{Non relativistic particle interacting with a time-independent magnetic field}
\label{Tweede}
\setcounter{equation}{0}
A particle of mass $m$ and charge $e$, with potential energy $\V$, moving in a Euclidean configuration space $\Q$, with cartesian coordinates $q^i$,  interacts with a (time-independent) magnetic field given by a closed two-form
$\F(q)=\frac{1}{2}\,F_{ij}(q)\,\dif q^i\wedge\dif q^j$. The dynamics is given by the Laplace equation:
\be\label{Laplace1}
m\,{\dif^2 q^i\over\dif t^2}=\delta^{ij}
\left(e\,F_{jk}(q)\,{\dif q^k\over\dif t}-
\frac{\partial \V(q)}{\partial q^j}\right)\,.\ee
Assuming $\Q$ to be Euclidean avoids topological subtleties, so that there exists a global potential one-form $\A(q)=A_i(q)\,\dif q^i$ such that $\F=\dif\A$. A global Lagrangian formalism can then be established with a Lagrangian function on the tangent bundle $\{\tau:T(\Q)\rightarrow\Q\}$:
\[\L(q,\dot{q})=\frac{1}{2}m\,\delta_{ij}\,\dot{q}^{\,i}\dot{q}^{\,j}+
e\,\dot{q}^{\,i}\,A_i(q)-\V(q)\,.\] 
The Euler-Lagrange equation is obtained as:
\bea
0&=&{\partial\L\over\partial q^i}-{\dif\over\dif t}{\partial\L\over\partial \dot{q}^{\,i}}=
-\frac{\partial \V}{\partial q^i}+
e\,\dot{q}^{\,k}{\partial A_k(q)\over\partial q^i}
-{\dif\over\dif t}\left(m\,\delta_{ij}\,\dot{q}^j+\,eA_i(q)\right)\nonumber\\
&=&
-\frac{\partial \V}{\partial q^i}+
e\,\dot{q}^{\,k}
\left({\partial A_k(q)\over\partial q^i}-{\partial A_i(q)\over\partial q^k}\right)
-m\,{\dif\over\dif t}\,\delta_{ij}\,\dot{q}^{\,j}\nonumber\\
\label{Laplace2}
&=&
-\frac{\partial \V}{\partial q^i}+
e\,\F_{ik}(q)\,\dot{q}^k
-m\,\delta_{ij}\,\ddot{q}^{\,j}\,,\eea
and coincides with the Laplace equation {\bf(\ref{Laplace1})}.\\
The Legendre transform
\[(q^i,\dot{q}^{\,j})\rightarrow\left(q^i,p_k={\partial\L\over\partial \dot{q}^{\,k}}=m\,\delta_{kl}\,\dot{q}^{\,l}+\,eA_k(q)\right)\,,\]
defines the Hamiltonian on the cotangent bundle $\{T^*(\Q)\stackrel{\kappa}{\rightarrow}\Q\}$:
\[\H_\A(q,p)=-\L(q,\dot q)+p_i\,\dot{q}^{\,i}={1\over 2m}\delta^{kl}(p_k-\,eA_k(q))\,
(p_l-\,eA_l(q))+\V(q)\,.\]
With the canonical symplectic two-form 
\be\label{omega1}
\omega_0=\dif q^i\wedge\dif p_i\,,\ee
the Hamiltonian vector field of $\H_\A$ is:
\[\X_\H
=\frac{\delta^{ij}}{m}(p_j-eA_j)\,{\partial\over\partial{\bf q}^i}
+\left(\frac{e}{m}\,\delta^{kl}\,{\partial A_k\over\partial q^i}(p_l-eA_l)\,
-\frac{\partial \V}{\partial q^i}\right)
{\partial\over\partial{\bf p}_i}\,.\]
Its integral curves are solutions of:
\be\label{Laplace3}
{\dif\,q^i\over\dif t}=\frac{\delta^{ij}}{m}(p_j-\,eA_j)\;,\;
{\dif\,p_i\over\dif t}=\frac{e}{m}\,\delta^{kl}\,{\partial A_k\over\partial q^i}(p_l-\,eA_l)\,
-\frac{\partial \V}{\partial q^i}\,,\ee
which is again equivalent to {\bf(\ref{Laplace1})}.\\
If the second de Rham cohomology were not trivial, $H^2_{dR}(Q)\not=0$, there is no
global potential $\A$ and a local Lagrangian formalism is needed. 
This can be done enlarging the configuration space $\Q$ to the total space ${\cal P}$ of a principal $U(1)$ bundle over $\Q$ with a connection, given locally by $\A$\footnote{See e.g. \cite{Baletal}}. This can be avoided using a global Hamiltonian formalism\footnote{Well known in symplectic mechanics, see e.g.{\bf\cite{Souriau,Abraham,Guillemin}}.} in the cotangent bundle $T^*(\Q)$ using a modified symplectic two-form:
\be\label{omega2}
\omega=\omega_0-e\F=\dif q^i\wedge\dif p_i-\frac{1}{2}e\,F_{ij}(q)\,\dif q^i\wedge\dif q^j\,,\ee
and a "charge-free" Hamiltonian: 
\[\H_0(p,q)=\frac{1}{2m}\,\delta^{kl}\,p_k\,p_l\,+\,\V(q)\,.\]
The Hamiltonian vector fields corresponding to an observable 
$f(q,p)$ are now defined relative to $\omega$ as
$\imath_{\X_f^\F}\,\omega=\dif f$ and given by:
\[\X_f^\F=
{\partial f\over\partial p_i}\,{\partial\over\partial{\bf q}^i}
-\left({\partial f\over\partial q^l}+{\partial f\over\partial p_k}eF_{kl}(q)\right)\,
{\partial\over\partial{\bf p}_l}\,.\]
With the Hamiltonian $\H_0$, the dynamics are again given by the Laplace equation {\bf(\ref{Laplace1})} in the form:
\be\label{Laplace4}
{\dif\,q^i\over\dif t}=\frac{\delta^{ij}}{m}p_j\;;\;{\dif\,p_l\over\dif t}=-\,
\delta^{ki}\left(\frac{\partial \V}{\partial q^i}+\frac{e}{m}p_i\,F_{kl}(q)\right)\,.\ee
The Poisson brackets, relative to the symplectic structure {\bf\ref{omega2}}, are: 
\be\label{Poisson3}
\bigl\{f,g\bigr\}=
{\partial f\over\partial q^i}
\,{\partial g\over\partial p_i}\,-
{\partial f\over\partial p_i}\,{\partial g\over\partial q^i}
+ {\partial f\over\partial p_k}\,e\,F_{kl}(q)\,{\partial g\over\partial p_l}
\;.\ee
In particular, the coordinates themselves have Poisson brackets:
\bea\label{Poisson4}
&&
\bigl\{q^i,q^j\bigr\}=0\;,\;\bigl\{q^i,p_l\bigr\}={\delta^i}_l\;,\nonumber\\
&&
\bigl\{p_k,q^j\bigr\}=-\,{\delta_k}^j\;,\;\bigl\{p_k,p_l\bigr\}=e\,F_{kl}(q)\;.\eea
Obviously, the meaning of the $\{q,p\}$ variables in {\bf(\ref{omega1})} and {\bf(\ref{omega2})} are different. However both formalisms $\left(\omega_0,\H_\A\right)$ and $\left(\omega,\H_0\right)$ lead to the same equations of motion and thus, they must be equivalent. Indeed, in each open set $U$ homeomorphic to ${\bf R}^6$, the vanishing $\dif \F=0$ implies the existence of $\A$ such that $\F=\dif\A$ in $U$ and, locally:
\[\omega=\dif q^i\wedge\dif p_i-\,\frac{1}{2}e\,F_{ij}\,\dif q^i\wedge\dif q^j=
-\,\dif[(p_i+e\,A_i)\,\dif q^i].\]
Thus there exist local Darboux coordinates:
\be\label{Darboux1}
\xi^i=q^i\;,\;\pi_k=p_k+e\,A_k(q)\;,\ee
such that $\omega=\dif\xi^i\wedge\dif\pi_i$, which is the form {\bf(\ref{omega1})}.\\
The dynamics defined by the Hamiltonian $\H_0(q,p)=p^2/2m+\V(q)$, with symplectic two-form 
$\omega$, is equivalent to the dynamics defined by  the Hamiltonian  
$\H_\A(\xi,\pi)=(\pi-e\,A(\xi))^2/2m+\V(\xi)$ and canonical symplectic structure $\omega=\dif \xi^i\wedge\dif \pi_i$. 
Equivalence is trivial since both symplectic two-forms are equal, but expressed in different coordinates $\{q,p\}$ and $\{\xi,\pi\}$, related by {\bf(\ref{Darboux1})}. It seems worthwhile to note that a gauge transformation $\A\rightarrow \A^\prime=\A+{\bf grad}\phi$ corresponds to a change of Darboux coordinates \[\{\xi^i,\pi_k\}\,\Rightarrow\,\{\xi^{i\,\prime}
=\xi^i,\pi^{\,\prime}_k=\pi_k+e\partial_k\phi\}\;,\]
i.e. a symplectic transformation.
\section{Noncommutative coordinates}
\label{Derde}
\setcounter{equation}{0}
Let us consider an affine configuration space $\Q=\A^N$ so that points of phase space, identified with $\M\doteq{\bf R}^{2N}={\bf R}_q^N\times{\bf R}_p^N$, may be given by linear coordinates $(q,p)$. Together with the (usual) magnetic field $\F$, we may introduce a (dual) magnetic field $\G=1/2\;G^{kl}(p)\,\dif p_k\wedge\dif p_l$, a closed two-form, $\dif\G=0$, in ${\bf R}_p^n$ space. Let $e$ be the usual electric charge and $r$, a dual charge, which couples the particle with $\F$ and $\G$. Consider the closed two-form:
\bea\label{omega3}
\omega&=&\omega_0-e\F+r\G\nonumber\\
&=&\dif q^i\wedge\dif p_i-\frac{1}{2}e\,F_{ij}(q)\,\dif q^i\wedge\dif q^j
+\frac{1}{2}r\,G^{kl}(p)\,\dif p_k\wedge\dif p_l\,.\eea
In matrix notation this two-form {\bf (\ref{omega3})} is represented as:
\bea\label{matrix}
\left(\Omega\right)&=&
\left(\begin{array}{cc}
-e\,\F & \Een \\
-\Een & +r\,\G
\end{array}\right)\nonumber\\
&&\nonumber\\
&=&
\left(\begin{array}{cc}
0 & \Een \\
\Een & +r\,\G
\end{array}\right)\,
\left(\begin{array}{cc}
-\Psi & 0 \\
0 & \Een
\end{array}\right)\,
\left(\begin{array}{cc}
\Een & 0 \\
-e\,\F & \Een
\end{array}\right)\nonumber\\
&&\nonumber\\
&=&
\left(\begin{array}{cc}
e\,\F & \Een \\
\Een & 0
\end{array}\right)\,
\left(\begin{array}{cc}
-\Een & 0 \\
0 & \Phi
\end{array}\right)\,
\left(\begin{array}{cc}
\Een & -r\,\G \\
0 & \Een
\end{array}\right)
\,.\eea
where\footnote{Observe that 
${\Phi_k}^\ell={\delta_k}^\ell-\,e\F_{kj}\,r\G^{j\ell}$ and 
${\Psi^i}_j={\delta^i}_j-\,r\G^{i\ell}\,e\F_{\ell j}$ are mutually transposed and that the matrices ${\Psi^k}_j\,r\,\G^{j\ell}=r\,\G^{kj}\,{\Phi_j}^\ell$ and 
${\Phi_k}^j\,e\,\F_{j\ell}=e\,\F_{kj}\,{\Psi^j}_\ell$ are antisymmetric.}
 $\Phi=(\Een-e\F\,r\G)\;;\;\Psi=(\Een-r\G\,e\F)$.
\\
The fundamental Hamiltonian equation $\imath_\X\omega=\dif f$, in {\bf(\ref{fundamenteel})}, reads:
\be\label{Fondamental1}
(X^i\,-r\,G^{ij}X_j)\,\dif p_i\,-\,(X_k-eF_{kl}X^l)\,\dif q^k
\,=\,\frac{\partial f}{\partial q^k}\,\dif q^k
\,+\,\frac{\partial f}{\partial p_i}\,\dif p_ i\;.\ee
This can be rewritten as
\be\label{Fondamental2}
(\frac{\partial f}{\partial p_i}\,-\,r\,G^{ij}\frac{\partial f}{\partial q^j})
={\Psi^i}_j\,X^j\;;\;
(\frac{\partial f}{\partial q^k}\,-\,e\,F_{kl}\frac{\partial f}{\partial p^l})
=-\,{\Phi_k}^l\,X_l\,.\ee
Obviously, from {\bf(\ref{matrix})} or {\bf(\ref{Fondamental2})}, the closed two-form $\omega$ will be non degenerate, and hence symplectic, if
${\bf det}(\Omega)={\bf det}(\Psi)={\bf det}(\Phi)\not=0$, so that $\left(\Omega\right)$ has an inverse:
\bea\label{inverse1}
\left(\Omega\right)^{-1}&=&
\left(\begin{array}{cc}
\Een & 0\\
+e\,\F & \Een 
\end{array}\right)\,
\left(\begin{array}{cc}
-\Psi^{-1} & 0 \\
0 & \Een 
\end{array}\right)\,
\left(\begin{array}{cc}
-r \G & \Een \\
\Een & 0 \end{array}\right)\nonumber\\
&=&
\left(\begin{array}{cc}
+\Psi^{-1}\,r\,\G & -\Psi^{-1}\\
+e\F\,\Psi^{-1}\,r\G+\Een & -e\F\Psi^{-1}
\end{array}\right)\;;\\
&&\nonumber\\
&=&
\left(\begin{array}{cc}
\Een & +r\G \\
0 & \Een 
\end{array}\right)\,
\left(\begin{array}{cc}
-\Een & 0 \\
0 & \Phi^{-1} 
\end{array}\right)\,
\left(\begin{array}{cc}
0 & \Een \\
\Een & -e\F \end{array}\right)\nonumber\\
\label{inverse2}
&=&
\left(\begin{array}{cc}
+r\,\G\,\Phi^{-1} & -r\G\,\Phi^{-1}\,e\F-\Een \\
\Phi^{-1}&-\Phi^{-1}\,e\F
\end{array}\right)\,.
\eea
Explicitely:
\be\label{omegaflat}
\omega^\flat:\dif f\rightarrow\left\{\begin{array}{ll}
(X_f)^i&=\;\;{(\Psi^{-1})^i}_j\,
\left(\partial f/\partial p_j\,-\,r\,G^{jk}\,\partial f/\partial q^k\right)\\
&\\
(X_f)_k&=-{(\Phi^{-1})_k}^l\,
\left(\partial f/\partial q^l\,-\,e\,F_{lj}\,\partial f/\partial p_j\right)
\end{array}
\right.\ee
The corresponding Poisson brackets are given by:
\be\label{Poisson4bis}
\left\{f,g\right\}=\omega(\X_f,\X_g)\,=\,
(\partial_q\,f\;\;\partial_p\,f)\;
\left(\Lambda\right)\;\left(\begin{array}{c}\partial_q\,g\\ \partial_p\,g\end{array}\right)\ee
with the matrix
\be\label{inverse3}\left(\Lambda\right)=-\left(\Omega\right)^{-1}=
\left(\begin{array}{cc}
-(\Psi^{-1}\,r\G=r\G\,\Phi^{-1}) & +\Psi^{-1}\\
-\Phi^{-1}&+(\Phi^{-1}\,e\F=e\F\,\Psi^{-1})
\end{array}\right)\,.
\ee
Explicitely:
\bea\label{Poisson5}
\left\{f,g\right\}
&=&
-{\partial f\over\partial q}(\Psi^{-1}\,rG){\partial g\over\partial q}
-{\partial f\over\partial p}(\Phi^{-1}){\partial g\over\partial q}\nonumber\\
&&
+{\partial f\over\partial q}(\Psi^{-1}){\partial g\over\partial p}
+{\partial f\over\partial p}(\Phi^{-1}\,eF){\partial g\over\partial p}\;.
\eea
%
In particular, for the coordinates $(q^i,p_k)$, we have:
\bea\label{Poisson6}
\bigl\{q^i,q^j\bigr\}&=&-{(\Psi^{-1})^i}_k\,rG^{kj}\,=\,
-\,rG^{ik}\,{(\Phi^{-1})_k}^j\,,\nonumber\\
\bigl\{q^i,p_l\bigr\}&=&{(\Psi^{-1})^i}_l\;,\nonumber\\
\bigl\{p_k,q^j\bigr\}&=&-\,{(\Phi^{-1})_k}^j\;,\nonumber\\
\bigl\{p_k,p_l\bigr\}&=&{(\Phi^{-1})_k}^j\,e\,F_{jl}\,=\,
e\,F_{kj}\,{(\Psi^{-1})^j}_l\;.\eea
With $\H(q,p)=(\delta^{kl}\,p_k\,p_l/2m)+\V(q)$, the equations of motion read: 
\bea\label{Hamilton}
\frac{dq^i}{dt}&=&
\left\{q^i,\H\right\}=
{(\Psi^{-1})^i}_j\,\left(-\,rG^{jk}\;\frac{\partial \H}{\partial q^k}
+\,\frac{\partial \H}{\partial p_j}
\right)\;,\nonumber\\
&=&
{(\Psi^{-1})^i}_j\,\left(-\,rG^{jk}\;\frac{\partial \V}{\partial q^k}+\,\frac{p^j}{m}
\right)\;,\nonumber\\
\frac{dp_k}{dt}&=&
\left\{p_k,\H\right\}=\,
{(\Phi^{-1})_k}^l\,\left(-\,\frac{\partial \H}{\partial q^l}
+\,eF_{lj}\,\frac{\partial \H}{\partial p_j}\right)\nonumber\\
&=&
{(\Phi^{-1})_k}^l\,\left(-\,\frac{\partial \V}{\partial q^l}
+\,eF_{lj}\,\frac{p^j}{m}\right)\;.\eea
The celebrated Darboux theorem guarantees the existence of local coordinates $(\xi^{\,i},\pi_k)$, such that $\omega=\dif \xi^i\wedge\dif \pi_i$. When one of the charges $(e,r)$ vanishes, such Darboux coordinates are easily obtained using the potential one-forms $\A=A_i(q)\dif q^i$ and $\widetilde{\A}=\widetilde{A}^k(p)\dif p_k$, such that $\F=\dif\A$ and $\G=\dif\widetilde{\A}$.\\ 
Indeed, if $r=0$, as in section {\bf\ref{Tweede}}, Darboux coordinates are provided by 
$\xi^i=q^i\,;\,\pi_k=p_k+e\,A_k(q)$.
A modified symplectic potential and two-form are defined by:
\be\label{potentiel1}
 \theta=(p_k+eA_k)\,\dif q^k\;;\;\omega=-\,\dif\theta\,.\ee
The Hamiltonian and corresponding equations of motion are:
\be\label{Hamiltoneen}
\H(\xi,\pi)={1\over 2}\delta^{kl}(\pi_k-eA_k(\xi))(\pi_l-eA_l(\xi))+\V(\xi)\,,\ee
\be\label{risnul}
{\dif\,\xi^i\over\dif t}=\delta^{ij}\,(\pi_j-eA_j(\xi))\;,\;
{\dif\,\pi_i\over\dif t}=e\,\delta^{kl}(\pi_k-\,eA_k)\,{\partial A_l\over\partial \xi^i}\,
-\frac{\partial \V}{\partial \xi^i}\,,\ee
which yields the second order equation in $\xi$, as in {\bf(\ref{Laplace1})}:
\be\label{Laplace5}
{\dif^2\,\xi^i\over\dif t^2}=\delta^{ij}\left(-\,\frac{\partial\V(\xi)}{\partial\xi^j}+\,eF_{jl}(\xi)\,
{\dif\,\xi^l\over\dif t}\right)\,.\ee
When $e=0$, Darboux variables are 
\be\label{Darboux2}
\xi^i=q^i+\,r\widetilde{A}^i(p)\,;\,\pi_k=p_k\;,\ee
and we define
\be\label{potentiel2}
\theta=p_k\,\dif( q^k+r\widetilde{A}^k)\;;\;\omega=-\,\dif\theta\,.\ee
The Hamiltonian and equations of motion are now given by:
\be\label{Hamiltontwee}
\H(\xi,\pi)={1\over 2}\delta^{kl}\,\pi_k\,\pi_l+\V(\xi-r\,\widetilde{A}(\pi))\,,\ee
\be\label{eisnul}
{\dif\,\xi^i\over\dif t}=\delta^{ij}\,\pi_j-\,r\partial_k\V(q)\,
\frac{\partial \widetilde{A}^k}{\partial \pi_i}\;,\;
{\dif\,\pi_i\over\dif t}=-\,\frac{\partial \V}{\partial q^i}(q)\,.\ee
The second order equation, obeyed by $\pi$ {\bf(!)}, is given by
\be\label{Laplace6}
{\dif^2\,\pi_i\over\dif t^2}=\,\partial^{\,2}_{ij}\V(q)\,
\left(-\,\delta^{jk}\pi_k+\,rG^{jk}(\pi)\,\frac{\dif \pi_l}{\dif t}\right)\,.\ee
Here the $q$-variable is assumed to be solved in terms of $\dot\pi$ from equation $\dot\pi_k=\,-\,\partial\V(q)/\partial q^k$ and this is possible if 
$\;{\bf det}(\partial^{\,2}_{ij}\V(q))\not=0$ !\\
In the case of nonzero charges $(e,r)$ and non-constant $\F$ and $\G$ fields, there is no generic formula to define global Darboux coordinates $(\xi^i,\pi_k)$. However, if 
the fields $\F$ and $\G$ are constant, the Poisson matrix {\bf (\ref{matrix})} is brought in 
canonical Darboux form by a linear symplectic orthogonalization procedure, \`{a} la Hilbert-Schmidt. 
In the next section this is done explicitely for $N=2$ and $N=3$. Obviously such a linear transformation:
$(q^i,p_k)\Rightarrow(\xi^i,\pi_k)$ is defined up to a linear symplectic map of ${\bf Sp}(2n)$. 
These variables $(\xi^i,\pi_k)\in{\bf R}^{2n}$ can be canonically quantised as operators obeying the commutation relations
\be\label{Quantizatie1}
\left[\widehat{\xi^i},\widehat{\xi^j}\right]=0\;;\;
\left[\widehat{\xi^i},\widehat{\pi_l}\right]=i\,\hbar\,{\delta^i}_l\;;\;
\left[\widehat{\pi_k},\widehat{\pi_l}\right]=0\,.\ee
As von Neumann taught us in \cite{John}, they are realised on the Hilbert space of square integrable functions of the variable $\xi$ as 
\be\label{Quantizatie2}
(\widehat{\xi^i}\Psi)(\xi)=\xi^i\,\Psi(\xi)\;;\;
(\widehat{\pi_k}\Psi)(\xi)=\frac{\hbar}{i}\,\frac{\partial\Psi(\xi)}{\partial\xi^k}\;.\ee
The original variables $(q^i,p_k)$ being linear functions of the $(\xi^i,\pi_k)$ are then also quantised.\\
When ${\bf det}(\Psi)={\bf det}(\Phi)=0$, the closed two-form $\omega$ is singular. When its rank is constant, $\omega$ defines a presymplectic structure on phase space which we
call the primary constraint manifold denoted by $\M_1$. The consistency of the resulting constrained Hamiltonian system will be examined in the $N=2$ and $N=3$ cases.
\section{Examples: $N=2$ and $3$}\label{Vierde}
\setcounter{equation}{0}
In the two examples below, we consider a classical Hamiltonian of the form 
\be\label{hamiltonien}
\H\,=\,\frac{1}{2m}\delta^{kl}\,p_k\,p_l\,+\,\V(q)\,.\ee 
A complete resolution will be given for a harmonic oscillator potential:
\be\label{erg}
\V(q)\doteq
\frac{\kappa}{2}\delta_{ij}\,q^i\,q^j\;.\ee
Also of interest is the case of a constant "electric field":
$\V(q)\,=\,-\,\E_k\,q^k$, which is exactly solvable and left to the reader.
\subsection{Dynamics in the noncommutative plane}
\label{vierdeA}
The magnetic fields in two dimensions, are written as:
\be\label{een}
e\,F_{ij}=B\,\epsilon_{ij}\;;\;r\,G^{kl}=C\,\epsilon^{kl}
\,,\ee
where $B$ and $C$ are pseudoscalars. The closed two-form {\bf(\ref{omega3})} becomes
\be\label{Nistwee}
\omega=
\dif q^i\wedge\dif p_i-B\,\dif q^1\wedge\dif q^2+C\,\dif p_1\wedge\dif p_2\,.\ee
The equation $\imath_X\omega=\dif f$ reads 
\be\label{vgl}
X^i-C\epsilon^{ij}X_j\,=\,\frac{\partial f}{\partial p_i}\;;\;
X_k-B\epsilon_{kl}X^l=-\frac{\partial f}{\partial q^k}\;.\ee
Denoting $\chi\doteq(1+C\,B)$, the matrices $\Phi$ and $\Psi$ are written as 
${\Phi_i}^j=\chi\,{\delta_i}^j$ and ${\Psi^k}_l=\chi\,{\delta^k}_l$. The matrix {\bf(\ref{matrix})} is then invertible if $\chi$ does not vanish. 
\subsubsection{The non degenerate case}\label{nietontaard}
Here, we will assume $\chi$ to be strictly positive.
The above equation {\bf(\ref{vgl})} can then be inverted with Hamiltonian vector fields given by:
\be\label{Fondamental4}
X^i=\chi^{-1}
\left(\frac{\partial f}{\partial p_i}-
C\,\epsilon^{ij}\frac{\partial f}{\partial q^j}\right)\;,\;
X_k=-\,\chi^{-1}
\left(\frac{\partial f}{\partial q^k}-
B\,\epsilon_{kl}\frac{\partial f}{\partial p^l}\right)
\,.\ee
The Poisson brackets {\bf(\ref{Poisson6})} become:
\bea\label{Poisson7}
\bigl\{q^i,q^j\bigr\}=-\,
C\,\chi^{-1}\epsilon^{ij}&;&
\bigl\{q^i,p_l\bigr\}=
\chi^{-1}\,{\delta^i}_l\;,\nonumber\\
\bigl\{p_k,q^j\bigr\}=
-\,\chi^{-1}\,{\delta_k}^j
&;&
\bigl\{p_k,p_l\bigr\}=
B\,\chi^{-1}\,\epsilon_{kl}
\;.\eea
Substitution of the Ansatz
\be\label{Ansatztwee}
\xi^i=
\alpha\,q^i+\beta\,\frac{C}{2}\,p_k\,\epsilon^{ki}\;;\;
\pi_k=
\gamma\,\frac{B}{2}q^j\,\epsilon_{jk}+\delta\,p_k\;,\ee
in the canonical Poison brackets, leads to the equations 
\bea\label{equations}
&&
\alpha^2-\alpha\beta-\frac{C\,B}{4}\beta^2=0\;,\;
\delta^2-\delta\gamma-\frac{C\,B}{4}\gamma^2=0\;,\nonumber\\
&&\alpha\delta+\frac{C\,B}{2}(\alpha\gamma+\delta\beta)-\frac{C\,B}{4}\beta\gamma=\chi\,.\eea
We choose the solution:
\be\label{solution}
\alpha=\delta=\sqrt{u}\;;\;\beta=\gamma=\frac{1}{\sqrt{u}}
\;;\;u=\frac{1}{2}(1+\sqrt{\chi})\;,\ee
such that {\bf(\ref{Ansatztwee})} reduces to {\bf(\ref{Darboux1})} when $C=0$  or to {\bf(\ref{Darboux2})} in case $B=0$.
The 2-form {\bf(\ref{omega3})} has the canonical Darboux form $\omega=d\xi^i\wedge d\pi_i$ in the variables 
\be\label{Darboux5}
\xi^i=
\sqrt{u}\left(q^i-\frac{C}{2u}\,\epsilon^{ik}\,p_k\right)\;;\;
\pi_k=
\sqrt{u}\left(p_k-\frac{B}{2u}\,\epsilon_{ki}\,q^i\right)\;.\ee
These have an inverse if, and only if $\chi\not=0$: 
\be\label{Darboux6}
\sqrt{\chi}\;q^i\,=\,
\sqrt{u}\left(\xi^i+\,\frac{C}{2u}\,\epsilon^{ik}\,\pi_k\right)\;;\;
\sqrt{\chi}\;p_k\,=\,\sqrt{u}\left(\pi_k+\,\frac{B}{2u}\,\epsilon_{ki}\,\xi^i\right)\,.
\ee
With the complex variables
\be\label{komplex1}
q=q^1+\im\,q^2\;,\;p=p_1+\im\,p_2\;;\;\xi=\xi^1+\im\,\xi^2\;,\;\pi=\pi_1+\im\,\pi_2\;,\ee
the above changes of variables are written as:
\be\label{komplex2}
\xi=\sqrt{u}\left(q+\im\,\frac{C}{2u}p\right)\;;\;
\pi=\sqrt{u}\left(p+\im\,\frac{B}{2u}q\right)\;.\ee
The inverse transformations are:
\be\label{komplex3}
q=\sqrt{u/\chi}\left(\xi-\im\frac{C}{2u}\pi\right)\;;\;
p=\sqrt{u/\chi}\left(\pi-\im\frac{B}{2u}\xi\right)\;.\ee
The Hamiltonian {\bf(\ref{erg})} becomes 
\bea\label{energie}
\H&=&\frac{1}{2m^{\,\prime}}\,\delta^{kl}\,\pi_k\,\pi_l\,
+\,\frac{\kappa^{\,\prime}}{2}\,\delta_{ij}\,\xi^i\,\xi^j
\,-\,\omega_L^{\,\prime}\,\Lambda\nonumber\\
&=&\frac{1}{2m^{\,\prime}}\,\frac{\pi^\dagger\pi+\pi\pi^\dagger}{2}
\,+\,
\frac{\kappa^{\,\prime}}{2}\,\frac{\xi^\dagger\xi+\xi\xi^\dagger}{2}
\,-\,
\omega_L^{\,\prime}\,\Lambda\;,\eea
where $\Lambda$ is angular momentum 
\bea\label{draaimoment}
\Lambda&=&\frac{1}{2}\left(\epsilon_{ij}\,\xi^i\,\delta^{jk}\pi_k-\epsilon^{kl}\,\pi_k\,\delta_{lj}\xi^j\right)\nonumber\\
&=&\frac{1}{2}\left((\xi^1\pi_2-\xi^2\pi_1)\,-\,(\pi_1\xi^2+\pi_2\xi^1)\right)\nonumber\\
&=&\frac{1}{4\im}\left((\xi^\dagger\pi-\xi\pi^\dagger)\,-\,(\pi\xi^\dagger+\pi^\dagger\xi)\right)\;.\eea
The "renormalised" mass and elasticity constant are given by:
\be\label{renorm}
\frac{1}{m^{\,\prime}}=
\frac{1}{m}\,\frac{u}{\chi}\,\left(1+\,\frac{c^2}{4\,u^2}\right)\;;\;
\kappa^{\,\prime}=
\kappa\,\frac{u}{\chi}\,\left(1+\,\frac{b^2}{4\,u^2}\right)\;.\ee
where 
\be\label{bepaling}
b=\frac{B}{\sqrt{m\kappa}}\;;\;c=C\sqrt{m\kappa}\,.\ee
The corresponding frequency $\omega^{\,\prime}_0=\sqrt{\kappa^{\,\prime}/m^{\,\prime}}$ is given 
in terms of the "bare" frequency $\omega_0=\sqrt{\kappa/m}$ by:
\be\label{frekwentie}
\omega^{\,\prime}_0\,=\,\frac{\omega_0}{2\chi}\left(\left(b-c\right)^2
\,+\,4\chi\right)^{1/2}\;.\ee
and $\omega_L^{\,\prime}$, the induced Larmor frequency, by:
\be\label{Larmor}
\omega_L^{\,\prime}=\frac{\omega_0}{2\chi}\left(b-c\right)\,.\ee
The solution of Hamiltonian's equations with {\bf(\ref{energie})} is standard. 
With\footnote{In the limit $\chi\rightarrow 0$, we have 
$\;m^\prime\omega^\prime_0=\sqrt{m^\prime\kappa^\prime}\rightarrow |B|$.} 
\be\label{B}
m^\prime\omega^\prime_0=\sqrt{m^\prime\kappa^\prime}=
\sqrt{m\kappa}\,
\left(\left(1+\frac{b^2}{4u^2}\right)
\left(1+\frac{c^2}{4u^2}\right)^{-1}\right)^{1/2}
\,\ee
reduced variables are introduced by:
\be\label{red}
Q\doteq (m^\prime\omega^\prime_0)^{1/2}\,\xi
\;;\;
P\doteq(m^\prime\omega^\prime_0)^{-1/2}\,\pi\;.\ee
The original $(q,p)$ are expressed as:
\bea\label{verander}
q&=&\sqrt{u/\chi}\,(m^\prime\omega^\prime_0)^{-1/2}
\left(Q-\im\,\frac{c^{\,\prime}}{2u}\,P\right)\,,\nonumber\\
p&=&\sqrt{u/\chi}\,(m^\prime\omega^\prime_0)^{+1/2}
\left(P-\im\,\frac{b^{\,\prime}}{2u}\,Q\right)\;,\eea
where
\be\label{bepalingbis}
c^{\,\prime}=C\,(m^\prime\omega^\prime_0)\,=\,C\,\sqrt{m^\prime\kappa^\prime}\;,\;
b^{\,\prime}=B/(m^\prime\omega^\prime_0)\,=\,B/\sqrt{m^\prime\kappa^\prime}\;.\ee
The symplectic structure and the Poisson brackets are:
\bea\label{complex2}
\omega&=&
\frac{1}{2}\left(\dif Q^\dagger\wedge\dif P\,+\,\dif Q\wedge\dif P^\dagger\right)\nonumber\\
\left\{f,g\right\}&=&2\left(
\frac{\partial f}{\partial Q}\frac{\partial g}{\partial P^\dagger}
+\frac{\partial f}{\partial Q^\dagger}\frac{\partial g}{\partial P}
-\frac{\partial f}{\partial P}\frac{\partial g}{\partial Q^\dagger}
-\frac{\partial f}{\partial P^\dagger}\frac{\partial g}{\partial Q}\right)\,.\eea
The fundamental nonzero Poisson bracket is 
\be\label{complex3}
\{Q,P^\dagger\}=2\;.\ee
In these variables, the Hamiltonian {\bf(\ref{energie})} reads: 
\be\label{ergter}
\H=\frac{\omega^{\,\prime}_0}{4}\,
\left((P^\dagger P+PP^\dagger)+(Q^\dagger Q+QQ^\dagger)\right)\,-\,
\omega^{\,\prime}_L\,\Lambda\;,\ee
where
\be\label{draaien}
\Lambda
=\frac{1}{4\im}\left((Q^\dagger P-QP^\dagger)\,-\,
(PQ^\dagger+P^\dagger Q)\right)\;.\ee
The corresponding equations of motion are:
\bea\label{beweging}
\frac{d Q}{dt}=\{Q,\H\}\,&=&\,2\,\frac{\partial \H}{\partial P^\dagger}\,=\,
\omega^{\,\prime}_0\,P\,-\,\im\,\omega_L^{\,\prime}\,Q\nonumber\\
\frac{d P}{dt}=\{Q,\H\}\,&=&-\,2\,\frac{\partial \H}{\partial Q^\dagger}\,=\,
-\,\omega^{\,\prime}_0\,Q\,-\,\im\,\omega_L^{\,\prime}\,P\;.\eea
With the shift variables
\be\label{shift1}
A_{(+)}=\frac{1}{2}\left(Q+\im\,P\right)\;;\;
A_{(-)}=\frac{1}{2}\left(Q^\dagger+\im\,P^\dagger\right)\,,\ee
the symplectic structure and the Poisson brackets are given by:
\be\label{shift2}
\omega\,=\,
-\im\left(\dif A_{(+)}^\dagger\wedge\dif A_{(+)}+
\dif A_{(-)}^\dagger\wedge\dif A_{(-)}\right)\,,
\ee
\bea\label{shift3}
\left\{f,g\right\}&=&-\im\left(
\frac{\partial f}{\partial A_{(+)}}\frac{\partial g}{\partial A_{(+)}^\dagger}
+\frac{\partial f}{\partial A_{(-)}}\frac{\partial g}{\partial A_{(-)}^\dagger}\right.
\nonumber\\
&&\quad\quad\left.
-\,\frac{\partial f}{\partial A_{(+)}^\dagger}\frac{\partial g}{\partial A_{(+)}}
\,-\,\frac{\partial f}{\partial A_{(-)}^\dagger}\frac{\partial g}{\partial A_{(-)}}\right)\;,
\eea
with fundamental nonzero brackets:
\be\label{shift4}
\{A_{(\pm)},A_{(\pm)}^\dagger\}=-\im\;.\ee
The Hamiltonian, with the (positive !) frequencies
\be\label{shift5}
\omega_{(\pm)}=(\omega^{\,\prime}_0\,\pm\,\omega^{\,\prime}_L)\,,\ee
reads now:
\be\label{shift6}
\H=\frac{\omega_{(+)}}{2}\,
\left(A_{(+)}^\dagger A_{(+)}+A_{(+)}A_{(+)}^\dagger\right)
\,+\,
\frac{\omega_{(-)}}{2}\,
\left(A_{(-)}^\dagger A_{(-)}+A_{(-)}A_{(-)}^\dagger\right)\;.
\ee
The corresponding equations of motion and their solutions are given by:
\be\label{shift7}
\frac{d A_{(\pm)}}{dt}\,=\,\{A_{(\pm)},\H\}\,=\,-\,\im\,\frac{\partial \H}{\partial A_{(\pm)}^\dagger}\,=\,-\,\im\,\omega_{(\pm)}\,A_{(\pm)}\;;\ee
\be\label{shift8}
A_{(\pm)}(t)={\bf exp}\left\{-\im\,\omega_{(\pm)}\,t\right\}\,
A_{(\pm)}(0)\,.\ee
The relations between variables are given by:
\bea\label{shift9}
A_{(+)}&=&\frac{1}{2}\left(Q+\im P\right)\nonumber\\
&=&
\frac{\sqrt{u}}{2}\left((m^\prime\omega^\prime_0)^{+1/2}\,(1-\frac{b^{\,\prime}}{2u})\,q
+\,\im\,(m^\prime\omega^\prime_0)^{-1/2}\,(1+\frac{c^{\,\prime}}{2u})\,p\right)\nonumber\\
A_{(-)}^\dagger&=&\frac{1}{2}\left(Q-\im P\right)\nonumber\\
&=&
\frac{\sqrt{u}}{2}\left((m^\prime\omega^\prime_0)^{+1/2}\,(1+\frac{b^{\,\prime}}{2u})\,q
-\,\im\,(m^\prime\omega^\prime_0)^{-1/2}\,(1-\frac{c^{\,\prime}}{2u})\,p\right)\,.
\eea
The inverse transformations are:
\bea\label{shift10}
q&=&(m^\prime\omega^\prime_0)^{-1/2}
\sqrt{u/\chi}\left(Q-\im\frac{c^{\,\prime}}{2u}P\right)\;,\nonumber\\
&=&(m^\prime\omega^\prime_0)^{-1/2}
\sqrt{u/\chi}\left((1-\frac{c^{\,\prime}}{2u})A_{(+)}
+(1+\frac{c^{\,\prime}}{2u})A_{(-)}^\dagger\right)\;,\nonumber\\
p&=&(m^\prime\omega^\prime_0)^{+1/2}
\sqrt{u/\chi}\left(P-\im\frac{b^{\,\prime}}{2u}Q\right)\nonumber\\
&=&\im\,(m^\prime\omega^\prime_0)^{+1/2}
\sqrt{u/\chi}\left((1-\frac{b^{\,\prime}}{2u})A_{(-)}^\dagger
-(1+\frac{b^{\,\prime}}{2u})A_{(+)}\right)\;.
\eea
Quantisation is trivial though the substitution of the fundamental Poison brackets 
{\bf(\ref{complex3}),(\ref{shift4})} by operator commutators 
\be\label{Kwantum1}
\left[{\bf Q},{\bf P}^\dagger\right]=2\im\,\hbar\;;\;
\left[{\bf A}_{(\pm)},{\bf A}_{(\pm)}^\dagger\right]=\hbar\,.\ee
Having kept the initial ordering, the quantum Hamiltonian has eigenvalues:
\be\label{Kwantum2}
E(n_{(+)},n_{(-)})=\hbar\omega_{(+)}\,(n_{(+)}+1/2)\,+\,\hbar\omega_{(-)}\,(n_{(-)}+1/2)\,,\ee
where $n_{(\pm)}$ are nonnegative integers. The corresponding eigenvectors are denoted by 
$|n_{(+)},n_{(-)}>$.
\subsubsection{The degenerate or constraint case}\label{ontaard}
The condition $\chi\doteq(1+BC)=0$ determines $\omega$ as a presymplectic structure on $\M$ and shall be called the primary constraint. Again, the notation is simplified using complex variables\footnote{Recall that with complex variables $q=q^1+\im\,q^2$, the diferentials $dq=dq^1+\im\,dq^2$ and $dq^\dagger=dq^1-\im\,dq^2$ have local dual vector fields $\{\partial/\partial q=(\partial/\partial q^1-\im\,\partial/\partial q^2)/2\;;\;\partial/\partial q^\dagger=(\partial/\partial q^1+\im\,\partial/\partial q^2)/2$ and similarly for the $p=p_1+\im\,p_2$ variables.}.
The presymplectic two-form reads
\bea\label{presymplectic}
\omega&=&\frac{1}{2}\left(dq^\dagger\wedge dp+dq\wedge dp^\dagger\right)\nonumber\\
&&
-\frac{B}{4\,\im}\left(dq^\dagger\wedge dq-dq\wedge dq^\dagger\right)
+\frac{C}{4\,\im}\left(dp^\dagger\wedge dp-dp\wedge dp^\dagger\right)\,.\eea
The Hamiltonian {\bf(\ref{erg})} becomes  
\be\label{ergske}
\H=\frac{1}{2m}\,\frac{p^\dagger p+p\,p^\dagger}{2}+
\frac{\kappa}{2}\,\frac{q^\dagger q+q\,q^\dagger}{2}\,,\ee
Writing a vector field as 
\[\X=X^i\,\partial/\partial q^i+X_k\,\partial/\partial p_k=U\,\partial/\partial q+U^\dagger\,\partial/\partial q^\dagger+V\,\partial/\partial p+V^\dagger\,\partial/\partial p^\dagger\;,\]
\bea\label{imath}
\imath_X\omega&=&\frac{1}{2}
\left((U+\im\,C\,V)\,dq^\dagger+(U^\dagger-\im\,C\,V^\dagger)\,dq\right.\nonumber\\
&&
\quad\left.-(V+\im\,B\,U)\,dp^\dagger-(V^\dagger-\im\,B\,U^\dagger)\,dp\right)\,.\eea
The homogeneous equation, $\imath_\Z\omega=0$ has nontrivial solutions. 
Indeed, with $U_0=Z^1+\im\,Z^2$ and $V_0=Z_1+\im\,Z_2$, equation {\bf(\ref{imath})} yields the system:
\be\label{Homogeen}
U_0+\im\,C\,V_0=0\;;\;\hbox{or}\quad V_0+\im\,B\,U_0=0\,,\ee
of which the determinant is $\chi=1+BC=0$.\\
The inhomogeneous equation $\imath_\X\omega\,=\,\dif \H$, i.e. the Hamiltonian dynamics, 
reads
\be\label{Inhomogeen}
U+\im\,C\,V=2\,\frac{\partial \H}{\partial p^\dagger}=\frac{p}{m}\;;\;
V+\im\,B\,U=-2\,\frac{\partial \H}{\partial q^\dagger}=\kappa\,q\,.\ee
It will have a solution if 
\be\label{secondary1}
\langle\dif\H|\Z\rangle=0\;.\ee 
This condition, termed secondary constraint, is explicitely given by:
\be\label{secondary2}
\frac{\partial\H}{\partial p}-\im\,C\,\frac{\partial\H}{\partial q}=0
\;;\;\hbox{or}\quad
\frac{\partial\H}{\partial q}-\im\,B\,\frac{\partial\H}{\partial p}=0\;.\ee
For the Hamiltonian {\bf(\ref{ergske})} 
this condition {\bf(\ref{secondary2})} is linear: 
\be\label{secondary3}
\frac{1}{m}\,p+\im\,C\,\kappa\,q=0\;;\;\hbox{or}\quad
\kappa\,q+\im\,B\,\frac{1}{m}\,p=0\,.\ee
and defines the secondary constraint manifold $\M_2$. \\
\underline{\bf On $\M_2$}, a particular solution of $\imath_\X\omega\,=\,\dif \H$ is given by:
\be\label{particuliere}
U_P\,=\,\frac{p}{m}\;;\;V_P\,=\,0\;.\ee
The general solution is given by:
\be\label{algemene}
U\,=\,\frac{p}{m}\,+\,U_0\;;\;
V\,=\,V_0\;.\ee
where $(U_0,V_0)$ is restricted to obey {\bf(\ref{Homogeen})}. 
This vector field, restricted to $\M_2$, should conserve the constraints i.e. must be tangent to $\M_2$: 
\be\label{rakend}
0=\langle\frac{1}{m}\,\dif p\,+\,\im\,C\,\kappa\,\dif q|X\rangle\;,\ee
The vector fields $U$ and $V$ are completely defined on $\M_2$, with ensuing equations of motion:
\bea\label{eindoplossing1}
\frac{dq}{dt}&=&U\,=\,-\im\frac{\sqrt{m\kappa}C}{1+m\,\kappa\,C^2}\,\omega_0\,q
\,=\,\frac{1}{1+m\,\kappa\,C^2}\,\frac{p}{m}\,,\nonumber\\
\frac{dp}{dt}&=&V\,=\,-\im\frac{\sqrt{m\kappa}C}{1+m\,\kappa\,C^2}\,\omega_0\,p
\,=\,-\frac{m\,\kappa\,C^2}{1+m\,\kappa\,C^2}\,\kappa\,q\;.\eea
In terms of the frequency:
\be\label{ontaardefre}
\omega_r=-\,\frac{\sqrt{m\kappa}\,C}{1+m\,\kappa\,C^2}\,\omega_0\,=\,
\frac{B/\sqrt{m\kappa}}{1+B^2/m\,\kappa}\,\omega_0\;,\ee
the solution is given by
\be\label{eindoplossing3}
q(t)=
{\bf exp}\left\{\im\,\omega_r\,t\right\}\,q_0\;;\;
p(t)=
{\bf exp}\left\{\im\,\omega_r\,t\right\}p_0\;.\ee
Obviously, if $q_0$ and $p_0$ obey the secondary constraints {\bf (\ref{secondary3})}, $q(t)$ and $p(t)$ obey them at all times.\\
The same result can be obtained by symplectic reduction, restricting the pre-symplectic two-form {\bf(\ref{presymplectic})} to $\M_2$:
\be\label{reduktie1}
\omega_{|\M_2}\,=\,-\im\,\frac{(1+m\kappa C^2)^2}{2C}\,dq^\dagger\wedge dq\,.\ee
\be\label{reduktie2}
\{f,g\}_{\M_2}\,=\,\frac{2\im\,C}{(1+m\kappa C^2)^2}
\left(\frac{\partial f}{\partial q^\dagger}
\frac{\partial g}{\partial q}\,-\,
\frac{\partial f}{\partial q}
\frac{\partial g}{\partial q^\dagger}\right)\,.\ee
The fundamental Poisson bracket is 
\be\label{reduktie}
\{q,q^\dagger\}_{\M_2}\,=\,\frac{-\,2\im\,C}{(1+m\kappa C^2)^2}\ee
The dynamics are given by:
\be\label{reduktie3}
\frac{dq}{dt}\,=\,-\,\frac{2\im\,C}{(1+m\kappa\,C)^2}\,\frac{\partial \H_r}{\partial q^\dagger}\,.
\ee
And, with the reduced Hamiltonian $\H_r$ given by 
\be\label{reduktie4}
\H_r\,=\,(1+m\kappa\,C^2)\,\frac{\kappa}{2}\,q^\dagger\,q\,,\ee
this yields equation {\bf(\ref{eindoplossing3})}. 
When $B>0$, hence $C<0$, we define 
\be\label{reduktie5}
a=\frac{(1+m\kappa\,C^2)}{|2\,C|}\,q^\dagger\,,\ee
such that
\be\label{reduktie6}
\{a,a^\dagger\}=-\,\im\;;\;
\H_r\,=\,\frac{\omega_r}{2}\,(a^\dagger\,a\;+\;a\,a^\dagger)\;.\ee
Quantisation is again trivial introducing operators ${\bf a}$ and ${\bf a}^\dagger$, obeying
\be\label{Kwantum3}
[{\bf a},{\bf a}^\dagger]\,=\,\hbar\,\ee
such that the quantum Hamiltonian 
\be\label{Kwantum4}
{\bf H}_r\,=\,\frac{\omega_r}{2}\,
({\bf a}^\dagger\,{\bf a}\;+\;{\bf a}\,{\bf a}^\dagger)\;.\ee
has eigenvalues:
\be\label{Kwantum5}
E(n)=\hbar\omega_r\,(n+1/2)\;.\ee
\subsubsection{The $\chi\rightarrow 0$ limit of {\bf(\ref{nietontaard})}.}
\label{benadering}
We need the expansion of 
\be\label{ben1}
(m^\prime\omega^\prime_0)=(m\,\omega_0)\times
\left(\left(1+\frac{b^2}{4u^2}\right)\left(1+\frac{c^2}{4u^2}\right)^{-1}\right)^{1/2}\,,\ee
in powers of $\epsilon=\sqrt{\chi}$, where $1+bc=\epsilon^2$ and $2u=1+\epsilon$. 
\bea\label{ben2}
(m^\prime\omega^\prime_0)&=&
\frac{m\,\omega_0}{|c|}\,
\left(1+\frac{c^2-1}{c^2+1}\,\epsilon+\cdots\right)\nonumber\\
&=&
\frac{1}{|C|}\,
\left(1+\frac{c^2-1}{c^2+1}\,\epsilon+\cdots\right)\nonumber\\
(m^\prime\omega^\prime_0)^{-1}&=&
\frac{(m\,\omega_0)^{-1}}{|b|}\,
\left(1+\frac{b^2-1}{b^2+1}\,\epsilon+\cdots\right)\nonumber\\
&=&
\frac{1}{|B|}\,
\left(1+\frac{b^2-1}{b^2+1}\,\epsilon+\cdots\right)
\,.\eea
Also, from {\bf(\ref{bepalingbis})}, we obtain
\bea\label{ben3}
\frac{c^{\,\prime}}{2u}&=&\frac{(m^\prime\omega^\prime_0)\,C}{2u}\,=\,
\frac{C}{|C|}\,\left(1-\frac{2}{c^2+1}\,\epsilon+\cdots\;,\right)\nonumber\\
\frac{b^{\,\prime}}{2u}&=&\frac{B}{(m^\prime\omega^\prime_0)\,2u}\,=\,
\frac{B}{|B|}\,\left(1-\frac{2}{b^2+1}\,\epsilon+\cdots\;,\right)\,.
\eea
For definitenees, we assume in the following $B>0$ and so $C<0$ in the limit $\epsilon\rightarrow 0$. We obtain
\bea\label{ben4}
1-\frac{b^{\,\prime}}{2u}=\frac{2}{1+b^2}\,\epsilon+\cdots\;&;&\;
1+\frac{b^{\,\prime}}{2u}=2-\frac{2}{1+b^2}\,\epsilon+\cdots\nonumber\\
1+\frac{c^{\,\prime}}{2u}=\frac{2}{1+c^2}\,\epsilon+\cdots\;&;&\;
1-\frac{c^{\,\prime}}{2u}=2-\frac{2}{1+b^2}\,\epsilon+\cdots\;.\eea
Also:
\be\label{ben5}
\omega^\prime_0=\frac{\omega_0}{2\epsilon^2}\,(b-c)
\left(1+\frac{2\epsilon^2}{(b-c)^2}\right)\;,\;
\omega^\prime_L=\frac{\omega_0}{2\epsilon^2}\,(b-c)\;,\ee
\bea\label{ben6}
\omega_{(+)}&=&\omega^\prime_0\,+\,\omega^\prime_L=
-\,\omega_0\,\frac{1+(m\omega_0)^2C^2}{(m\omega_0)C}\;\frac{1}{\epsilon^2}\,,\nonumber\\
\omega_{(-)}&=&\omega^\prime_0\,-\,\omega^\prime_L=
-\,\omega_0\,\frac{(m\omega_0)\,C}{1+(m\omega_0)^2\,C^2}\,.\eea
One of the frequencies $\omega_{(+)}$ diverges, while the other $\omega_{(-)}$ tends to 
$\omega_r$ defined in {\bf(\ref{ontaardefre})}.
The relations in {\bf(\ref{shift9})} yield the initial conditions:
\bea\label{ben7}
A_{(+)}(0)
&\approx&
\sqrt{\frac{|B|}{2}}\,(1+b^2)^{-1}\left(q_0+\im\,\frac{B}{(m\omega_0)^2}\;p_0\right)\,(\epsilon+O(\epsilon^2))
\nonumber\\
A_{(-)}^\dagger(0)
&\approx&
\sqrt{\frac{|B|}{2}}\,\left(q_0-\im\,\frac{1}{|B|}\,p_0\right)\,(1+O(\epsilon^2))\;.
\eea
The solutions {\bf(\ref{shift10})}, in the $\epsilon\rightarrow 0$ limit are then written as
\bea\label{ben8}
q(t)&\approx&
\sqrt{\frac{2}{|B|}}\,\left(\frac{1}{\epsilon}\,A_{(+)}(0)\,{\bf exp}\{-\im\,\omega_{(+)}t\}
\,+\,\frac{1}{1+c^2}\,A_{(-)}^\dagger(0)\,{\bf exp}\{\im\,\omega_rt\}\right)\nonumber\\
&\approx&
(1+b^2)^{-1}\,\left(q_0+\im\,\frac{|B|}{(m\omega_0)^2}\,p_0\right)
\,{\bf exp}\{-\im\,\omega_{(+)}t\}\nonumber\\
&&\,+\,
(1+c^2)^{-1}\,\left(q_0-\im\,|B|^{-1}\,p_)\right)\,{\bf exp}\{+\im\,\omega_rt\}\,;
\eea
The first term is a fast oscillating function with diverging frequency and so averages to zero. Furthermore, if the initial conditions are on $\M_2$, i.e. if $\left(q_0+\im\,|B|\,p_0/(m\omega_0)^2\right)=0$, this first term behaves as 
$O(\epsilon)\,{\bf exp}\{\im\,\nu\,t/\epsilon^2\}$ converging to zero. The second term is then reduced to the expression {\bf(\ref{eindoplossing3})} of $q(t)$. Similar considerations hold for $p(t)$ in such a way that the solution stays on $M_2$.
\subsection{Noncommutative ${\bf R}^{\bf 3}$}
In ${\bf R}^{\bf 3}$, the magnetic fields $\F$ and $\G$ are written in terms of pseudovectors $\B=\{B^k\}$ and $\C=\{C_k\}$ as: 
\be\label{drie}
e\,F_{ij}=\epsilon_{ijk}\,B^k\;;\;r\,G^{ij}=\epsilon^{ijk}\,C_k\,.\ee
The closed two-form {\bf(\ref{omega3})} is written as:
\be\label{Rdrie1}
\omega=
\dif q^i\wedge\dif p_i-\frac{1}{2}\epsilon_{ijk}B^k\,\dif q^i\wedge\dif q^j
+\frac{1}{2}\epsilon^{klm}C_m\,\dif p_k\wedge\dif p_l\,.\ee
The fundamental equation $\imath_X\omega=\dif f$ reads
\be\label{Rdrie2}
X^i-C_k\epsilon^{ijk}X_j\,=\,\frac{\partial f}{\partial p_i}\;;\;
X_k-B^i\epsilon_{kli}X^l=-\frac{\partial f}{\partial q^k}\;.\ee
Defining $\vartheta=\C\cdot\B=C_k\,B^k$ and $\chi=1+\vartheta$, this is also written as 
\bea\label{Rdrie3}
\chi\,X^i&=&
({\delta^i}_j+B^i\,C_j)\frac{\partial f}{\partial p_j}
-C_k\,\epsilon^{ijk}\frac{\partial f}{\partial q^j}\,\nonumber\\
\chi\,X_k&=&
-\left(({\delta_k}^l+C_k\,B^l)\frac{\partial f}{\partial q^l}
-B^i\,\epsilon_{kli}\frac{\partial f}{\partial p_l}\right)\,.
\eea
The $3\times 3$ matrices $\Phi$ and $\Psi$ read:
\[{\Phi_i}^j=\chi\,{\delta_i}^j-C_i\,B^j\;;
\;{\Psi^k}_l=\chi\,{\delta^k}_l-B^k\,C_l\;,\]
with $\det\Phi=\det\Psi=\chi^2$.
Assuming again $\chi\not=0$\footnote{The $(N=3)$ case wil only be examined in the nondegenerate case $\chi>0$.}, these matrices have inverses:
\[
{(\Phi^{-1})_i}^j=\frac{1}{\chi}\left({\delta_i}^j+C_i\,B^j\right)\;,\;
{(\Psi^{-1})^k}_\ell=\frac{1}{\chi}\left({\delta^k}_\ell+B^k\,C_\ell\right)\;.\]
The Hamiltonian vector fields are obtained from {\bf(\ref{Rdrie3})}:
\bea\label{Rdrie4}
X^i&=&\chi^{-1}\left(
({\delta^i}_j+B^i\,C_j)\frac{\partial f}{\partial p_j}
-C_k\,\epsilon^{ijk}\frac{\partial f}{\partial q^j}\right)\,,\nonumber\\
X_k&=&
-\,\chi^{-1}\left(
({\delta_k}^l+C_k\,B^l)\frac{\partial f}{\partial q^l}
-B^i\,\epsilon_{kli}\frac{\partial f}{\partial p_l}\right)\,.
\eea
The Poissson brackets are given by:
\bea\label{Rdrie5}
\bigl\{q^i,q^j\bigr\}=-
\chi^{-1}\;\epsilon^{ijk}\,C_k
\;&,&\;
\bigl\{q^i,p_l\bigr\}=
\chi^{-1}\,\left({\delta^i}_l+B^i\,C_l\right)\;,\nonumber\\
\bigl\{p_k,q^j\bigr\}=
-\,\chi^{-1}\,\left({\delta_k}^j+C_k\,B^j\right)
\;&,&\;
\bigl\{p_k,p_l\bigr\}=
\chi^{-1}\;\epsilon_{klm}\,B^m
\;.\eea 
The Ansatz {\bf (\ref{Ansatztwee})} has to be generalised to 
\bea\label{Ansatzdrie}
\xi^{i}&=&
\alpha\,q^i+\alpha^\prime\,B^i\,(C_k\,q^k)
-\beta\,\frac{1}{2}\epsilon^{ijk}\,p_j\,C_k\,;
\nonumber\\
\pi_k&=&
\alpha\,p_k+\alpha^{\,\prime}\,(p_i\,B^i)\,C_k+\beta\,\frac{1}{2}\epsilon_{klm}\,B^l\,q^m\;.
\eea
For $\alpha,\beta$ similar equations as in {\bf(\ref{equations})} are obtained:
\be\label{Rdrie6}
\alpha^2-\alpha\beta-\frac{\vartheta}{4}\beta^2=0\;,\;
\alpha^2+\vartheta(\alpha\beta)-\frac{\vartheta}{4}\beta^2=\chi\,,\ee
with a the same solution ($\chi$ assumed to be strictly positive):
\be\label{Rdrie7}
\alpha=\sqrt{u}\;;\;\beta=\frac{1}{\sqrt{u}}
\;;\;u=\frac{1}{2}(1+\sqrt{\chi})\;.\ee
Furthermore, there is an additional equation for $\alpha^\prime$:
\be\label{Rdrie8}
\chi\left(\vartheta\,{\alpha^\prime}^2+2\alpha\alpha^\prime\right)
+\left(\alpha^2-\alpha\beta+\frac{1}{4}\beta^2\right)=0\;.\ee
Substituting {\bf(\ref{Rdrie7})}, one obtains
\[\vartheta\,\alpha^{\prime 2}+2\sqrt{u}\alpha^\prime+\frac{1}{4u}=0\;,\]
with solution, remaining finite when $\vartheta\rightarrow 0$,:
\be\label{Rdrie9}
\alpha^\prime=\sqrt{u}\,\gamma=\frac{(1-\sqrt{u})}{\vartheta}\,.\ee
The formulae {\bf(\ref{Ansatzdrie})} are finally written as:
\bea\label{Rdrie10}
\xi^{i}&=&\sqrt{u}
\left(
q^i+\gamma\,B^i\,(C_k\,q^k)-\frac{1}{2u}\epsilon^{ijk}\,p_j\,C_k
\right)\,;
\nonumber\\
\pi_k&=&\sqrt{u}
\left(
p_k+\gamma\,(p_i\,B^i)\,C_k+\frac{1}{2u}\epsilon_{klm}\,B^l\,q^m
\right)\;.
\eea
In old fashioned vector notation, this appears as:
\bea\label{Rdrie11}
\ksi&=&\sqrt{u}
\left(
\q+\gamma\,\B\,(\C\cdot\q)-\frac{1}{2u}\,\p\times\C
\right)\,;
\nonumber\\
\pie&=&\sqrt{u}
\left(
\p+\gamma\,(\p\cdot\B)\,\C+\frac{1}{2u}\,\B\times\q\right)\;.
\eea
The inverse formulae of {\bf(\ref{Rdrie10})} are obtained as:
\bea\label{Rdrie12}
q^{i}&=&\frac{\sqrt{u}}{\sqrt{\chi}}
\left(
\xi^i+\gamma^{\,\prime}\,B^i\,(C_k\,\xi^k)
+\frac{1}{2u}\epsilon^{ijk}\,\pi_j\,C_k\right)\,;
\nonumber\\
p_k&=&\frac{\sqrt{u}}{\sqrt{\chi}}
\left(
\pi_k+\gamma^{\,\prime}\,C_k(\pi_l\,B^l)
-\frac{1}{2u}\epsilon_{klm}\,B^l\,\xi^m\right)\;,\eea
Or, in vector notation:
\bea\label{Rdrie13}
\q&=&\frac{\sqrt{u}}{\sqrt{\chi}}
\left(
\ksi+\gamma^{\,\prime}\,\B\,(\C\cdot\ksi)
+\frac{1}{2u}\,\pie\times\C\right)\,;
\nonumber\\
\p&=&\frac{\sqrt{u}}{\sqrt{\chi}}
\left(
\pie+\gamma^{\,\prime}\,\C\,(\pie\cdot\B)
-\frac{1}{2u}\,\B\times\ksi\right)\;,\eea
where 
\be\label{Rdrie14}
\gamma^{\,\prime}=\frac{\sqrt{\chi}-\sqrt{u}}{\vartheta\,\sqrt{u}}\,.\ee
Again, for sake of simplicity, we consider a configuration space which is Euclidean $\Q=\E^3$ with metric 
$<\overline{\bf v};\overline{\bf w}>=\delta_{ij}\,v^i\,w^j
=(\underline{\bf v}\cdot\overline{\bf w})$ such that $v_i=\delta_{ij}\,v^i$. 
Substitution of {\bf(\ref{Rdrie12})} in a Hamiltonian of the form {\bf(\ref{erg})}, leads  
to a Hamiltonian quadratic in $(\xi,\pi)$ and to a system of linear evolution equations. In the case when $\B$ and $\C$ point in the same direction:
\be\label{Rdrie15}
\B=B\,\overline{\bf e}_Z\;;\;\C=C\,\underline{\bf e}_Z\;,\ee
a particularly simple Hamiltonian is obtained. Parallel coordinates are defined by $\xi^3\,,\,\pi_3$ and transverse coordinate vectors
by $\ksi_{\perp}=\ksi-\,\xi^3\,\overline{\bf e}_Z$ and $\pie_{\perp}=\pie-\,\pi_3\,\underline{\bf e}_Z$.
Indeed, eq. {\bf(\ref{Rdrie12})} becomes
\bea\label{Rdrie16}
q^{1}=\frac{\sqrt{u}}{\sqrt{\chi}}
\left(\xi^1+\frac{1}{2u}\,\pi_2\,C\right)&,&
p_1=\frac{\sqrt{u}}{\sqrt{\chi}}
\left(\pi_1+\frac{1}{2u}\,\xi^2\,B\right)\;,\nonumber\\
q^{2}=\frac{\sqrt{u}}{\sqrt{\chi}}
\left(\xi^2-\frac{1}{2u}\,\pi_1\,C\right)&,&
p_2=\frac{\sqrt{u}}{\sqrt{\chi}}
\left(\pi_2-\frac{1}{2u}\,B\,\xi^1\right)\;,\nonumber\\
q^3=\xi^3&,&p_3=\pi_3\;.\eea
The Hamiltonian is:
\be\label{Rdrie17}
\H(\xi,\pi)=
\left(
\frac{1}{2\,m_{\perp}}\,(\pie_{\perp})^2 +\frac{k_{\perp}}{2}\,(\ksi_{\perp})^2
\right)
+\left(\frac{1}{2\,m}\,(\pi_3)^2 +\frac{k}{2}\,(\xi^3)^2\right)
+\H_{int}(\xi,\pi)\;.\ee
The transverse degrees of freedom are seen to have a renormalised\footnote{Due to  $\kappa^2+\kappa^{\,\prime\,2}(rC)^2\,(eB)^2+2\kappa\,\kappa^{\,\prime}\,rCeB=1$, the mass and elastic constant of the $z$ degrees of freedom, as expected, are not renormalised.} mass and elasticity constant which are given by the same expressions as in {\bf(\ref{renorm})}:
\be\label{Rdrie18}
\frac{1}{m_{\perp}}=
=\frac{1}{m}\,\frac{u}{\chi}\,\left(1+\,\frac{c^2}{4\,u^2}\right)\;;\;
\kappa_{\perp}=
\kappa\,\frac{u}{\chi}\,\left(1+\,\frac{b^2}{4\,u^2}\right)\;,\ee
where 
\[b=\frac{B}{\sqrt{m\kappa}}\;;\;c=C\sqrt{m\kappa}\,.\]
The fields $\B$ and $\C$ induce a sort of magnetic moment interaction along the $Z$-axis with the same Larmor frequency as before:
\be\label{Rdrie19}
\widetilde{\H}_{ind}(\xi,\pi)=-\omega_L^{\,\prime}\,\Lambda_3\;,\ee
where $\Lambda_3=\xi^1\,\pi_2-\xi^2\,\pi_1$.
Acrtually, the condition {\bf(\ref{Rdrie15})} reduces the $(N=3)$ case to a sum $(N=2)\oplus(N=1)$. 
The three relevant frequencies of our oscilator are:
\be\label{Rdrie20}
\omega_3=\sqrt{k/m}\;;\;\omega_\perp=\sqrt{k_\perp/m_\perp}\;;\;
\omega_L^{\,\prime}=\frac{1}{\chi}\omega_0\left(b-c\right)\,.\ee
The spectrum of the quantum Hamiltonian is easily obtained as 
\be\label{Rdrie21}
E(n_{(+)},n_{(-)},n_3)=\hbar\omega_{(+)}\,(n_{(+)}+1/2)\,+\,\hbar\omega_{(-)}\,(n_{(-)}+1/2)\,+\,\hbar\omega_3(n_3+1/2)\,,\ee
where $n_{(\pm)},n_3$ are nonnegative integers. Corresponding eigenvectors are denoted by 
$|n_{(+)},n_{(-)},n_3>$.
\section{Symmetries}\label{Vijfde}
For Euclidean configuration space $\Q\equiv\E^N$, with metric $\delta_{ij}$, an infinitesimal rotation is written as:
\be\label{draaien1}
\varphi:
q^i\rightarrow q^{\,\prime i}=
q^i+\frac{1}{2}\,\delta\epsilon^{\alpha\beta}{\left(M_{\alpha\beta}\right)^i}_j\,q^j\;,\ee
where ${\left(M_{\alpha\beta}\right)^i}_j={\delta^i}_\alpha\,\delta_{\beta j}-
{\delta^i}_\beta\,\delta_{\alpha j}$ are the generators of the rotation group obeying the Lie algebra relations:
\be\label{rotaties1}
[M_{\alpha\beta},M_{\mu\nu}]=-\,\delta_{\alpha\mu}\,M_{\beta\nu}
+\delta_{\alpha\nu}\,M_{\beta\mu}
-\delta_{\beta\nu}\,M_{\alpha\mu}
+\delta_{\beta\mu}\,M_{\alpha\nu}\,.\ee
This induces the push forward in $T^*(\Q)$:
\bea\label{draaien2}
&&
\widetilde{\varphi}:T^*(\Q)\rightarrow T^*(\Q):
(q^i,p_k)\rightarrow(q^{\,\prime i},{p^{\,\prime}}_k)\,,\nonumber\\
&&
q^{\,\prime\,i}=
q^i+\frac{1}{2}\,\delta\epsilon^{\alpha\beta}{\left(M_{\alpha\beta}\right)^i}_j\,q^j\;;\;
{p^{\,\prime}}_k=p_k-\frac{1}{2}\,\delta\epsilon^{\alpha\beta}\,
p_l{\left(M_{\alpha\beta}\right)^l}_k\,.\eea
In a basis\footnote{with dual basis $\{{\bf e}^{\alpha\beta}\}$ in $\L^*(SO(N))$.} $\{{\bf e}_{\alpha\beta}\}$ of $\L(SO(N))$, let ${\bf u}=(1/2){\bf e}_{\alpha\beta}\,u^{\alpha\beta}$ denote a generic element. With 
${\cal R}({\bf u})=\exp\left\{\frac{1}{2}\,u^{\alpha\beta}\,M_{\alpha\beta}\right\}$, finite rotations are written as
\be\label{draaien3}
q^i\rightarrow q^{\,\prime i}
={{{\cal R}({\bf u})}^i}_j\,q^j\;;\;
p_k\rightarrow {p^{\,\prime}}_k=p_l\,{{{\cal R}^{-1}({\bf u})}^l}_k\,.\ee
The vector field $\X_{\bf u}$ 
(see appendix {\bf\ref{mechanica}}) is given by its components:
\be\label{klein}
(X_{\bf u})^i=\frac{1}{2}\,u^{\alpha\beta}{(M_{\alpha\beta})^i}_j\,q^j\;;\;
(X_{\bf u})_k=-\frac{1}{2}\,u^{\alpha\beta}p_l{(M_{\alpha\beta})^l}_k\;.\ee
It conserves the canonical symplectic potential and two-form:
\[\L_{X_{\bf u}}\theta_0=0\;;\;\L_{X_{\bf u}}\omega_0=0\,.\]
The action is in fact Hamiltonian for the {\it canonical symplectic structure}. With the notation of appendix {\bf\ref{mechanica}}, we have 
\bea\label{draaimoment1}
&&\X_{\bf u}=\omega_0^\sharp(\dif\,\Xi({\bf u}))\;,\nonumber\\ 
&&\Xi({\bf u})=\frac{1}{2}\,u^{\alpha\beta}\,\J^0_{\alpha\beta}(q,p)\;,\nonumber\\
&&\J^0:T^*(\Q)\rightarrow\L^*(SO(N)):(q,p)\rightarrow\frac{1}{2}\J^0_{\alpha\beta}(q,p)\,
{\bf e}^{\alpha\beta}\,,\nonumber\\
&&\J^0_{\alpha\beta}(q,p)=p_k\,{(M_{\alpha\beta})^k}_j\,q^j\;.\eea
In terms of the momenta $\J^0_{\alpha\beta}$, the rotation {\bf(\ref{draaien1})} reads 
\be\label{draaimoment2}
\delta q^i=\frac{1}{2}\,\delta\epsilon^{\alpha\beta}\,\{q^i,\J^0_{\alpha\beta}\}_0\;;\;
\delta p_k=\frac{1}{2}\,\delta\epsilon^{\alpha\beta}\,\{p_k,\J^0_{\alpha\beta}\}_0\,.\ee
The Lie algebra relations {\bf(\ref{rotaties1})} become Poisson brackets:
\be\label{rotaties2}
\left\{\J^0_{\alpha\beta},\J^0_{\mu\nu}\right\}_0=
-\,\delta_{\alpha\mu}\,\J^0_{\beta\nu}
+\delta_{\alpha\nu}\,\J^0_{\beta\mu}
-\delta_{\beta\nu}\,\J^0_{\alpha\mu}
+\delta_{\beta\mu}\,\J^0_{\alpha\nu}\,.\ee
Naturally, for the modified symplectic structure {\bf(\ref{omega3})}, the action 
{\bf(\ref{draaien2})} will be symplectic  if, and only if, the magnetic fields obey:
\bea\label{voorwaarden}
F_{kl}(q)&=&F_{ij}({\cal R}({\bf u})\,q)\,{({\cal R}({\bf u}))^i}_k\,{({\cal R}({\bf u}))^j}_l\,,\\
G^{kl}(p)&=&\,{({\cal R}^{-1}({\bf u}))^k}_i\,{({\cal R}^{-1}({\bf u}))^l}_j\,
G^{ij}(p\,{\cal R}^{-1}({\bf u}))\,.\eea
For constant magnetic fields, this holds if ${\cal R}({\bf u})$ belongs to the intersection of the isotropy groups of $\F$ and $\G$, which, in three dimensions, is not empty if both magnetic fields are along the same axis. A rotation along this "z-axis" is then symplectic. However, in general it will not be Hamiltonian and there will be no momentum $\J_Z$ such that $\delta q=\{q,\J_Z\}$. Again the discussion simplifies when one of the charges $r$ or $e$ vanishes. If the potentials $\A$ or $\widetilde{\A}$ are invariant under ${\cal R}({\bf u})$, then the action is Hamiltonian\footnote{Exercise 4.2A in {\bf\cite{Abraham}}, defining a (generalized) Poincar\'{e} momentum.} with momentum defined by the symplectic potentials {\bf(\ref{potentiel1})} or {\bf(\ref{potentiel2})} as
\be\label{actie}
 \langle\J(q,p)|{\bf u}\rangle=\langle\theta_{(e,0)}|X_{\bf u}\rangle\;\hbox{or}\quad
\langle\theta_{(0,r)}|X_{\bf u}\rangle\;.\ee
Obviously there is always an $SO(N)$ group action on the $(\xi,\pi)$ coordinates which is Hamiltonian with respect to {\bf(\ref{omega3})} and momentum given by:
\be\label{draaimoment3}
 \J_{\alpha\beta}(\xi,\pi)=\pi_k\,{(M_{\alpha\beta})^k}_j\,\xi^j\;.
 \ee
However,the hamiltonian {\bf(\ref{erg})}, looking apparently $SO(N)$ symmetric, is explicitely seen not to be so when expressed in the $(\xi,\pi)$ variables. 
\section{Final Comments}\label{Zesde}
The symplectic structure in cotangent space, $T^*(\Q)\stackrel{\kappa}{\rightarrow} \Q$, was modified through the introduction of a closed two-form $\F$ on $T^*\Q$, which has the geometic meaning of the pull-back of the magnetic field $F$, a closed two-form on $\Q$: $\F=\kappa^*(F)$. A first caveat warns us that the other closed two-form $\G$ does not have such an intrinsic interpretation. Indeed, it is obvious that a mere change of coordinates in $\Q$ will spoil the form {\bf (\ref{omega3})} of $\omega$. This means that our approach must be restricted to configuration spaces with additional properties, which have to be conserved by coordinate changes. The most simple example is a flat linear\footnote{Quantum mechanics on a noncommutative shere $S^2$ and on general noncommutative Riemann surfaces was examined in 
{\bf (\cite{Nair,Morariu}}.}space $\Q=\E^N$, when {\bf (\ref{omega3})} is assumed to hold in linear coordinates. Obviously, a linear change in coordinates will then conserve this particular form.
Although the restriction to constant fields $\F$ and $\G$ is a severe limitation\footnote{In the case $e=0$, Darboux coordinates are given by {\bf (\ref{Darboux2})} and in {\bf \cite{Berard}} such model was considered with the possibility of having a monopole in $p$-space!}, it allowed us to find explicit Darboux coordinates {\bf (\ref{Ansatztwee})} when $N=2$ and {\bf (\ref{Ansatzdrie})} when $N=3$.\\
Finally, when $\det\{\Een-\,r\G\,e\F\}=0$, the closed two-form $\omega$ is degenerate with constant rank and defines a pre-symplectic structure on $T^*(\Q)$. Its null-foliation decomposes $T^*(\Q)$ in disjoint leaves and on the space of leaves, $\omega$ projects to a unique symplectic two-form. In two dimensions, the representations of the corresponding quantum algebra in Hilbert space and its reduction in the degeneracy case were studied in {\bf\cite{Duval,Nair,Morariu,Horvathy,PeterMikhail}}.
\appendix
\section
{Essential Symplectic Mechanics}
\label{mechanica}
\setcounter{equation}{0}
Let $\{\M,\omega\}$ be a symplectic manifold with symplectic structure defined by a two-form $\omega$ which is closed, $\dif\omega=0$, and nondegenerate such that the induced mapping $\omega^\flat:T(\M)\rightarrow T^*(\M):\X\rightarrow\imath_\X\omega$ has an inverse $\omega^\sharp:T^*(\M)\rightarrow T(\M):\alpha\rightarrow\omega^\sharp(\alpha)$.
The paradigm of a (non-compact) symplectic manifold is a cotangent bundle $T^*(\Q)$ of a differential configuration space $\Q$. 
In a coordinate system $\{q^i\}$ of $\Q$, a cotangent vector may be written as 
$\alpha_q=p_i\,\dif q^i$. This defines coordinates $z\Rightarrow\{q^i,p_k\}$ of points $z\in\M\equiv T^*(\Q)$ and an associated holonomic basis $\{\dif p_k,\dif q^i\}$ of $T_z^*(\M)$.
The canonical one-form is defined as $\theta_0\doteq p_i\,\dif q^i$. 
Obviously, the exact two-form $\omega_0\doteq-\,\dif\theta_0=\dif q^i\wedge\dif p_i$ is symplectic.
\\
To each observable, which is a differentiable function $f$ on $\{\M,\omega\}$, the symplectic structure associates a {\it Hamiltonian vector field}: 
\be\label{fundamenteel}
\X_f\doteq\omega^\sharp(\dif f)\quad\hbox{or}\quad\imath_{\X_f}\omega=\dif f\;.\ee
Such a vector field generates a one-parameter (local) transformation group:
$\T_f(t):\M\rightarrow\M:z_0\rightarrow z(t)$, solution of  
$\dif z(t)/\dif t=\X_f(z(t))\;,\;z(0)=z_0$. \\
In particular, {\it the} Hamiltonian $\H$ generates the dynamics of the associated mechanical system. With the usual interpretation of time, $\X_\H$ is assumed to be complete such that its  flux is defined for all $t\in[-\infty,+\infty]$. Transformations, induced by an Hamiltonian vector field $\X_f$, conserve the symplectic structure\footnote{${\T_f(t)}^*$ denotes the pull-back of ${\T_f(t)}$ and $\L$ is the Lie derivative along $\X_f$.}: 
\be\label{behouden}
{\T_f(t)}^*\omega=\omega\;\hbox{or, locally:}\;\L_{\X_f}\omega=0\,.\ee
More generally, the transformations conserving the symplectic structure form the group $Sympl(\M)$ of {\it symplectomorphisms} or {\it canonical transformations}. 
Vector fields obeying $\L_\X\omega=0$, generate canonical transformations and are called 
{\it locally Hamiltonian}, since \footnote{We use $\L_\X=\dif\,\imath_\X +\imath_\X\,\dif$ on differential forms.} $\dif\,\imath_{\X}\omega=0$ implies that, locally in some $U\subset\M$, there exists a function $f$ such that $\dif f_{|U}=(\imath_\X\omega)_{|U}$.\\
The {\it Darboux theorem} guarantees the existence of local charts $U\subset\M$ with 
coordinates $\{q^i,p_k\}$ such that, in each $U$, $\omega$ is written as:
\be\label{omega}
\omega_{|U}=\dif q^i\wedge \dif p_i\;.\ee
In the natural basis $\{{\bf \partial/\partial q^i},{\bf \partial/\partial p_k}\}$ of $T_z(\M)$, the Hamiltonian vector fields corresponding to $f$ reads
\[\X_f={\partial f\over\partial p_i}\,{\partial\over\partial{\bf q}^i}
-{\partial f\over\partial q^i}\,{\partial\over\partial{\bf p}_i}\,.\]
The {\it Poisson bracket} of two observables is defined by:
$\bigl\{f,g\bigr\}\doteq\omega(\X_f,\X_g)$, with the following properties:
\bean
&&\bigl\{f_1,f_2\bigr\}=-\bigl\{f_2,f_1\bigr\}\\
&&\bigl\{f_1,g_1\cdot g_2\bigr\}=\bigl\{f,g_1\bigr\}\cdot g_2+g_1\cdot\bigl\{f,g_2\bigr\}\\
&&\bigl\{f,\bigl\{g_1,g_2\bigr\}\bigr\}=\bigl\{\bigl\{f,g_1\bigr\},g_2\bigr\}+
\bigl\{g_1,\bigl\{f,g_2\bigr\}\bigr\}
\eean
These properties, relating the pointwise product $g_1\cdot g_2$ with the bracket $\bigl\{f,g\bigr\}$, are said to endow the set of differentiable functions on $\M$ with the structure of a {\it Poisson algebra} $\P(\M)$.
In a coordinate system $(z^A)$, where 
$\omega=\frac{1}{2}\,\omega_{AB}\,\dif z^A\wedge\dif z^B$, it is given by:
\be\label{Poisson0}\bigl\{f,g\bigr\}=
{\partial f\over\partial z^A}\,\Lambda^{AB}\,{\partial g\over\partial z^B}\;,\ee
where $\Lambda$ is minus $\omega^{-1}$. In Darboux coordinates it reads:
\be\label{Poisson1}
\bigl\{f,g\bigr\}_0=
{\partial f\over\partial q^i}
\,{\partial g\over\partial p_i}\,-
{\partial f\over\partial p_i}\,{\partial g\over\partial q^i}\;.\ee
The Poisson brackets of the Darboux coordinates themselves are:
\be\label{Poisson2}
\bigl\{q^i,q^j\bigr\}_0=0\;,\;\bigl\{q^i,p_l\bigr\}_0={\delta^i}_l\;,\;
\bigl\{p_k,q^j\bigr\}_0=-\,{\delta_k}^j\;,\;\bigl\{p_k,p_l\bigr\}_0=0\;.\ee
The dynamical evolution of an observable is given by:
\be\label{dynamica}
{\dif f\over \dif t}={\stackrel{\longrightarrow}{\X}}_\H(f)=
\imath_{\X_\H}\dif f=\imath_{\X_\H}\imath_{\X_f}\omega=\omega(\X_f,\X_\H)
=\bigl\{f,\H\bigr\}\,.\ee
%
A Lie group $G$ acts as a symmety group on a symplectic manifold $\M$, if there is a group homomorphism $\T:G\rightarrow Sympl(\M):g\rightarrow\T(g)$. An infinitesimal action defined by a Lie algebra element ${\bf u}\in\LG$ is given by the locally Hamiltonian vector field 
\be\label{klein0}
\X_{\bf u}(z)=\frac{d}{dt}\left(\T(\exp(t{\bf u}))z\right)_{|\,t=0}\,.\ee
When each $\X_{\bf u}$ is Hamiltonian, the group action is said to be {\it almost Hamiltonian} and $\bigl\{\M,\omega\bigr\}$ is called a {\it symplectic G-space}. In such a case, a linear map 
$\Xi:\LG\rightarrow\P(\M):{\bf u}\rightarrow\Xi({\bf u})$ can always be constructed such that
$\X_{\bf u}=\omega^\sharp(\dif\,\Xi({\bf u}))$. 
When there is a $\Xi$ which is also a Lie algebra homomorphism:
$\Xi([{\bf u},{\bf v}])=\bigl\{\Xi({\bf u}),\Xi({\bf v})\bigr\}$, the group is said to have a {\it Hamiltonian action} and $\bigl\{\M,\omega,\Xi\bigr\}$ is called a {\it Hamiltonian G-space}. Since $\Xi$ is linear in $\LG$, it defines a {\it momentum mapping} $\J$ from $\M$ to the dual $\LG^*$ of the Lie algebra defined by:
 $\langle\J(z)|{\bf u}\rangle=\Xi({\bf u},z)$. When $\M$ is a Hamiltonian $G$-space, the momentum mapping is equivariant under the action of $G$ on $\M$ and its co-adjoint action on $\LG^*$.\\
In general there may be topological obstructions to such a Lie algebra homomorphism. However, when $G$ acts on $\Q$:
$\varphi:G\rightarrow Diff(\Q):g\rightarrow\varphi(g):q\rightarrow q^{\,\prime}=\varphi(g)q$, the action is extended to a symplectic action in $\{\M=T^*(\Q),\omega_0\}$:
$\widetilde{\varphi}:G\rightarrow Sympl(\M):g\rightarrow\widetilde{\varphi}(g):(q,p)\rightarrow(q^{\,\prime},p^{\,\prime})$, 
 where $p^{\,\prime}$ is defined by $p=(\varphi(g))_{|q}^*\,p^{\,\prime}$. It follows that
$\widetilde{\varphi}(g)^*\theta_0=\theta_0\;;\;
\widetilde{\varphi}(g)^*\omega_0=\omega_0$.
The infinitesimal action is given by 
$\X_{\bf u}(z)=(\dif\widetilde{\varphi}(\exp(t{\bf u}))z/dt)_{|\,t=0}$
and $\L_{\X_{\bf u}}\theta_0=0\;;\;\L_{\X_{\bf u}}\omega_0=0$. From $\omega_0^\flat(\X_{\bf u})=\dif\langle\theta_0|\X_{\bf u}\rangle$, it follows that the action is almost Hamiltonian with $\Xi({\bf u})=\langle\theta_0|\X_{\bf u}\rangle$. Moreover, since
$\langle\theta_0|\X_{[{\bf u},{\bf v}]}\rangle=\omega_0(\X_{\bf u},\X_{\bf v})=\{\Xi({\bf u}),\Xi({\bf v})\}$, the action is Hamiltonian and $\bigl\{T^*(\Q),\omega_0,\Xi\bigr\}$ is a Hamiltonian $G$-space.
\section
{Presymplectic Mechanics}
\label{link}
\setcounter{equation}{0}
A manifold $\M_1$, endowed with a closed but degenerate\footnote{$\M_1$ is the primary constrained manifold, arising e.g. from a degenerate Lagrangian.} 2-form $\omega$, with constant rank, is said to be presymplectic. The mapping $\omega^\flat$ has a nonvanishing kernel, given by those nonzero vector fields $\X_0$ obeying $\omega^\flat(\X_0)\doteq\imath_{\X_0}\omega=0$. The fundamental dynamical equation 
\be\label{link1}
\omega^\flat(\X)=\dif \H\;,\ee
has then a solution if 
\be\label{link2}\langle \dif\H|\X_0\rangle\,
=\,0\quad;\;\forall\X_0\in{\cal K}er(\omega^\flat)\,.\ee
If this is nowhere satisfied on $\M_1$, the hamiltonian $\H$ does not define any dynamics on $\M_1$. When {\bf(\ref{link2})} is identically satisfied, a particular solution $\X_P$ of {\bf(\ref{link1})} is defined in the entire manifold $\M_1$ and so is the general solution obtained summing the general solution of the homogeneous equation $\imath_{\X_0}\omega=0$, i.e \(\X_G=\X_P+\X_0\), which will contain arbitrary functions.
When {\bf(\ref{link2})} is satisfied for some points $z\in \M_1$, we shall asssume they form a submanifold, called the secondary constrained submanifold with injection $\jmath_2:\M_2\hookrightarrow\M_1$. The particular solution $\X_P$ of {\bf(\ref{link1})} is now defined in $\M_2$ and so is the general solution $\X_G$. Requiring that $\X_G$ conserves the constraints amounts to ask that 
$\X_G$ is tangent to $\M_2$:
\be\label{link4}
\X_G=\jmath_{2\star}(\X_2)\;;\;\X_2\in\Gamma(\M_2,T\M_2)\,.\ee
 Again, when there are no points where this tangency condition is satisfied, {\bf(\ref{link1})} is meaningless. Another possibility is that some of the arbitrary functions in $\X_0$ become determined and the tangency condition is obeyed on the entire $\M_2$. The general solution then still contains some arbitrary functions. Finally it may happen that the conditions {\bf(\ref{link4})} are only satisfied on some $\M_3$ with $\jmath_3:\M_3\hookrightarrow\M_2$.
The story then goes on until one of the first two alternatives are reached.


\end{document}